\documentclass[twocolumn,amsmath,amssymb,aps,pre,reprint]{revtex4}
\usepackage{graphicx}
\begin{document}

\title{
Active Microrheology, Hall Effect, and Jamming in Chiral Fluids  
} 
\author{
C. Reichhardt and C.J.O. Reichhardt 
} 
\affiliation{
Theoretical Division and Center for Nonlinear Studies,
Los Alamos National Laboratory, Los Alamos, New Mexico 87545, USA
} 

\date{\today}
\begin{abstract}
We examine the motion of a probe particle driven through a chiral fluid
composed of circularly swimming disks.
We find that the probe particle travels in both the longitudinal direction,
parallel to the driving force, and in the transverse direction,
perpendicular to the driving force,
giving rise to a Hall angle.
Under constant driving force,
we show
that the probe particle velocity in both the longitudinal and transverse
directions
exhibits 
nonmonotonic behavior
as a function of
the activity of the circle swimmers.
The Hall angle is maximized when a resonance occurs
between the frequency 
of the chiral disks
and the
motion of the probe particle.
As the density of the chiral fluid increases,
the Hall angle gradually decreases before reaching zero
when the system enters a jammed state.
We show that the onset of jamming depends on the chiral particle
swimming frequency, with a fluid state appearing at low frequencies and
a jammed solid occurring at high frequencies.
\end{abstract}
\maketitle

\section{Introduction}

A variety of systems can be described
as assemblies of particles
that exhibit chiral or circular motion 
\cite{1,2}, such as
circularly moving colloids \cite{3,4,5,6},  
biological circle swimmers \cite{7}, active spinners \cite{8,9,10,11,12}, 
circularly driven particles \cite{13,14,15,N}, 
and chiral robot swarms \cite{16}.
Other systems in which chiral or gyroscopic motion occurs include
skyrmions in chiral magnets \cite{17,18}
and classical charged particles moving in a magnetic field \cite{19}. 
Such chiral particle assemblies can exhibit a variety of dynamical phases such
as large scale
rotations \cite{16}, self-assembly \cite{5,6,8,9,11},
edge currents \cite{5,9,15},
and odd viscosity responses \cite{10}. 

Viscosity, fluctuations, and jamming in particle assemblies can be examined
at the local level using
active rheology, which is based on the response of a probe particle
that is driven at either constant force or constant velocity
through a fluid or jammed medium 
\cite{20,21,22,23,24,25}. 
Active rheology has been
applied to the onset of 
jamming  \cite{21,22,26,27,28,29}, where the threshold for probe particle motion
increases from zero to a finite value at the jamming transition.
It has been used to measure
changes in viscosity and diffusive responses 
\cite{21,30,31,32,33,34,35}
as well as velocity-force relations \cite{20,21,27,36,37,38,39}.
Active rheology has been applied not only to soft matter systems, but also to the
dynamics of individual vortices dragged 
across pinning landscapes in type-II superconductors \cite{40,41,42}.
In systems that are active rather than passive, active rheology shows
large changes in the velocity of the probe particle 
as a function of increasing
bath activity when the system transitions from a fluid state to an actively
phase separated state \cite{43}.
In each case,
when the probe particle is driven at constant force, it moves in the direction
of drive and exhibits symmetric fluctuations in the transverse direction, with no
transverse drift or Hall velocity.

Here we study the active rheology
of a probe particle driven through a chiral fluid of circularly swimming disks.
We find that for low and intermediate fluid densities,
the probe particle
exhibits a longitudinal velocity $\langle V_{\rm long}\rangle$ in the direction of drive
as well as a finite
transverse or Hall velocity $\langle V_{\rm trans}\rangle$,
giving rise to a 
Hall effect with a Hall angle of
$\theta_{\rm Hall} = \arctan(\langle V_{\rm trans}\rangle/\langle V_{\rm long}\rangle)$.
We examine the evolution of the Hall angle as a
function of applied driving force, temperature, and 
chiral fluid density, and find
Hall angles that are as large as
$\theta_{\rm Hall} = 45^{\circ}$.
We also observe
non-monotonic behavior
of $\theta_{\rm Hall}$
in which the transverse velocity is maximized when a commensuration occurs between 
the chiral disk rotation frequency and the time interval between consecutive
collisions of the probe particle with the chiral disks.
In general, $\theta_{\rm Hall}$ decreases with increasing chiral disk density,
and it drops to zero at high densities when a jammed state appears.
In the dense limit, the probe particle can move only when the driving force is
larger than a finite threshold value, and this threshold depends strongly on the
chiral disk swimming frequency.
At low frequencies, the threshold is nearly zero,
while at high frequencies, the threshold increases when the system
acts like a solid.
We compare the dynamics of the probe particle to driven skyrmions 
which have recently been shown to exhibit a skyrmion Hall effect that also
exhibits nonmonotonic behavior as
a function of dc drive, temperature, and
skyrmion density \cite{44,45,46,47,48}.   

\section{Simulation and System}

We consider a two-dimensional $L \times L$
system with $L=36$
in which we place
$N$ non-overlapping disks with a radius $R_{d}=0.5$,
where the disks have repulsive harmonic interactions.  
The density of the system is characterized by the area covered by the 
disks, $\rho = N\pi R^2_{d}/L^2$.
For monodisperse disks at $T = 0$,
when $\rho=0.9$ the system forms a triangular solid
in which the disks are just touching.
The force between disks $i$ and $j$ is given by
${\bf F}^{ij}_{pp} = k(r_{ij} -2R_{d})\Theta(r_{ij} -2R_{d}){\bf \hat r}_{ij}$,
where $r_{ij} = |{\bf r}_{i} - {\bf r}_{j}|$,
$ {\bf \hat r}_{ij} = ({\bf r}_{i} - {\bf r}_{j})/r_{ij}$,
and $\Theta$ is the Heaviside step function.
The spring stiffness is set to $k = 50$, ensuring that the maximum
overlap between disks is less than one percent.
The densities and parameters we consider here
have also been studied in previous works \cite{15,27,43}.
The dynamics of disk $i$ is
determined by the following overdamped equation of motion:
\begin{equation}
\eta \frac{d{\bf r}_i}{dt} = \sum_{j \neq i}^{N}{\bf F}_{pp}^{ij} + {\bf F}_{\rm circ}^i + {\bf F}^{T}_{i}.  
\end{equation}
We set
the damping constant $\eta=1$ and
our simulation time step is 
$\Delta t = 0.002$.
Here
${\bf F}_{\rm circ}^i$ is a driving force
that creates a circular motion of the disks 
of the form  
${\bf F}_{\rm circ}^i = A(\sin(\omega t){\bf x}+\cos(\omega t){\bf y})$, 
controlled by
varying the drive amplitude $A$.
All of the chiral disks move in phase with each other.
The thermal force $F^{T}$ is produced by
Langevin kicks with the properties
$\langle F^{T}_{i}(t)\rangle = 0$ and
$\langle F^{T}_i(t)F^{T}_j(t^{\prime})\rangle = 2\eta k_BT\delta_{ij}\delta(t - t^{\prime})$.
Unless otherwise noted, we fix
$F^{T} = 2.0$, a value large enough to maintain the system in a liquid 
state up to the solidification
density $\rho=0.9$.
To create our probe particle, we select 
a single disk and replace ${\bf F}^i_{\rm circ}$ with 
${\bf F}_{D}=F_D{\bf {\hat x}}$.
We measure the average velocity response in the
longitudinal direction,
$\langle V_{\rm long}\rangle=\sum_i^T {\bf v}_p(t_i) \cdot {\bf \hat x}$,
as well as
in the
transverse direction, 
$\langle V_{\rm trans}\rangle=\sum_i^T {\bf v}_p(t_i) \cdot {\bf \hat y}$,
where ${\bf v}_p(t_i)$ is the instantaneous velocity of the probe particle.
 These quantities are averaged over an interval of
 $T=5\times 10^6$ time steps, which is long enough to ensure that
 the system has reached a steady dynamical state for the parameters we consider.
 In the absence of collisions between the probe particle and the chiral disks,
 we obtain the free flow value
 $\langle V_{\rm long}\rangle = F_{D}/\eta$.

\section{Results} 
\begin{figure}
\includegraphics[width=\columnwidth]{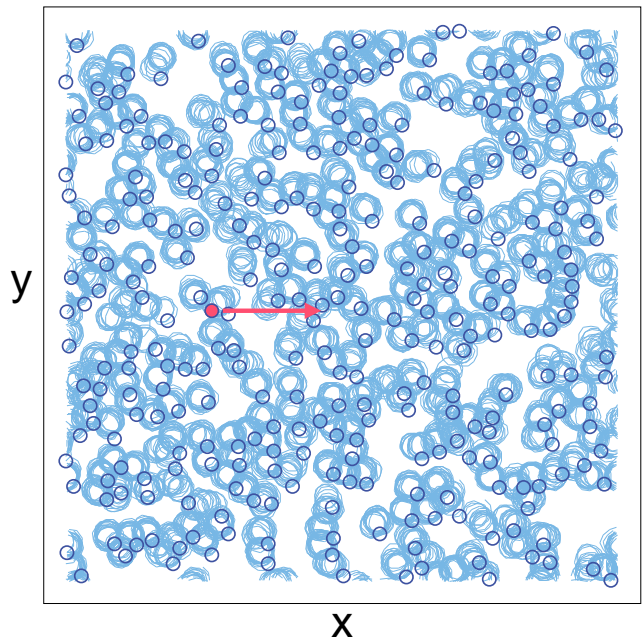}
\caption{Instantaneous positions (dark blue circles) and trajectories (light blue lines)
  of chiral
  disks during a fixed period of time along with the position (red circle) and driving
  direction (red arrow) of the probe particle
  in a sample with
  $\rho = 0.181$, $A = 2.5$, $\omega = 0.006$, $F^T=2.0$,
  and $F_{D} = 2.0$.
  The chiral disks undergo
a combination of diffusion and circular motion. 
}
\label{fig:1}
\end{figure}

In Fig.~\ref{fig:1} we show an image of the system
highlighting the chiral disk locations and trajectories
over a fixed period of time
for a system with $\rho = 0.181$, $A = 2.5$, $\omega = 0.006$,
and a thermal force of $F^{T} = 2.0$.
The disks execute circular orbits, and the center of mass of each circular
orbit has a diffusive behavior.
The red disk is the probe particle, which does not experience a circular drive but
instead
moves under a force
$F_{D}$ applied in the $x$ direction, as indicated by the arrow.

\begin{figure}
\includegraphics[width=\columnwidth]{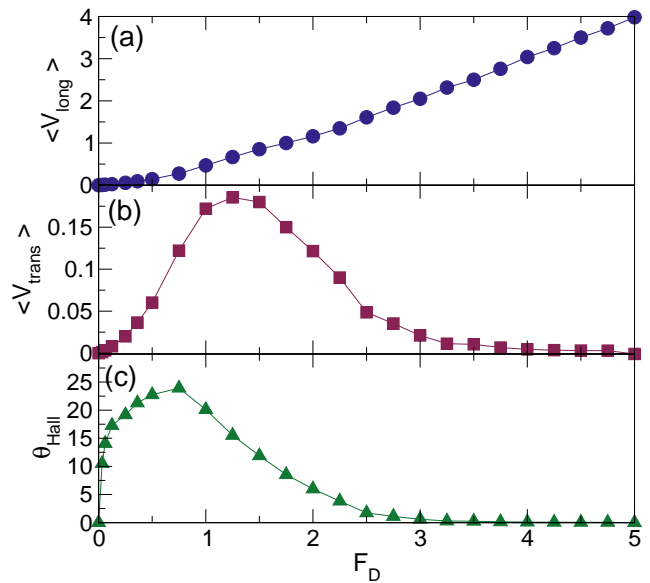}
\caption{Local probe response in a system with
  $\phi = 0.424$, $A = 2.5$ and $\omega = 0.006$. 
  (a) The longitudinal velocity $\langle V_{\rm long}\rangle$ vs $F_{D}$.
  (b) The transverse velocity $\langle V_{\rm trans}\rangle$ vs $F_{D}$.
  (c)
The Hall angle $\theta_{{\rm Hall}} = \arctan(\langle V_{\rm trans}\rangle/\langle V_{\rm long}\rangle)$  vs $F_D$.
}
\label{fig:2}
\end{figure}

In Fig.~\ref{fig:2}(a,b) we plot
$\langle V_{\rm long}\rangle$ and
$\langle V_{\rm trans}\rangle$, respectively, versus $F_D$ in
a system
with $\rho = 0.424$, $A = 2.5$, and $\omega = 0.006$.
Here $\langle V_{\rm long}\rangle$ monotonically increases  
with increasing $F_{D}$ and there is no
threshold for motion, while
$\langle V_{\rm trans}\rangle$
increases with
increasing drive at low $F_D$ before
reaching a maximum near $F_{D} = 1.25$  
and then decreasing again.
Since both the longitudinal and transverse velocities are finite,
the driven particle is moving at 
an angle with respect to drive direction,
similar to the Hall effect 
found for the motion of a charged particle in a magnetic field.
We plot
the Hall angle
$\theta_{\rm Hall} = \arctan(\langle V_{\rm trans}\rangle/\langle V_{\rm long}\rangle)$
versus $F_D$ in Fig.~\ref{fig:2}(c).
The maximum value of $\theta_{\rm Hall} = 23^{\circ}$
occurs at $F_{D} = 0.75$, a drive smaller than the value of $F_D=1.26$ at which 
the maximum in $\langle V_{\rm trans}\rangle$ appears. For higher drives,
$\theta_{\rm Hall}$ gradually 
deceases, reaching
a value close to zero for $F_{D} > 4.0$.  
  
\begin{figure}
\includegraphics[width=\columnwidth]{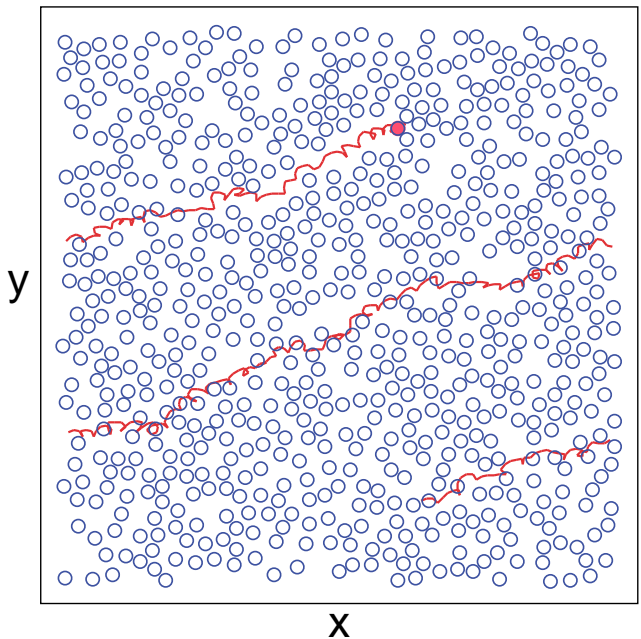}
\caption{Instantaneous positions of chiral disks (dark blue circles) and probe particle
  (red circle) along with the probe particle trajectory (red line) over a period of time
  for the
  system in Fig.~\ref{fig:2}
  with $\phi=0.424$, $A=2.5$, and $\omega=0.006$
  at $F_{D} = 1.0$,
  where the probe particle moves at an average Hall angle of $\theta_{\rm Hall} = 20^\circ$.  
}
\label{fig:3}
\end{figure}

In Fig.~\ref{fig:3} we illustrate the trajectory of the probe particle
over a fixed time interval superimposed on
a snapshot of the instantaneous chiral disk locations
for the system in Fig.~\ref{fig:2} at $F_{D} = 1.0$.
The probe particle is  moving at an angle 
of approximately $\theta_{\rm Hall}=20^\circ$ with respect to the drive;
however, there are local trajectory segments in which the Hall angle is larger or
smaller than average.

The chiral disks
have an intrinsic rotation frequency of $\omega$,
and therefore the time required for each chiral disk to complete one orbit is
$\tau_P = 2\pi/\omega$.
The average spacing between chiral disks is $a=1/\sqrt{\rho}$.
When the probe particle comes into
contact with a chiral disk at small $F_D$,
the chiral disk can complete multiple rotations during the
time required for the probe particle to move out of interaction range since
$F_D\tau_P \ll a$.
As a result, the average transverse force exerted on the probe
particle by the chiral disk is small and $\theta_{\rm Hall}$ is nearly zero.
At high $F_{D}$, the probe particle is moving rapidly in the longitudinal direction 
and spends a very short time interacting with the
chiral disks during a collision since
$F_D\tau_p \gg a$,
so once again the
maximum transverse
shift experienced by the probe particle is small and the Hall angle is small.
Between these limits, a resonance can occur.
When $\omega = 0.006$ and $\rho=0.424$, as in Fig.~\ref{fig:2},
the average spacing between chiral disks is
$a = 1.35$, and
$\tau_P = 1047 \Delta t$, where $\Delta t=0.002$ is the size of a simulation time step.
At $F_{D} = 0.75$, the probe particle would move a distance
$F_D\tau_p=1.57$
during one chiral rotation period in the absence of collisions with the chiral disks.
Collisions reduce this travel distance to a value that is
close to $a$,
so that on average the probe particle interacts with a given chiral disk
for one rotation period.
This maximizes the chance that the chiral disk will exert a coordinated, monodirectional
transverse force on the probe particle, resulting in a significant transverse displacement.
The maximum in $\theta_{\rm Hall}$ thus arises due to a
resonance
between the chiral rotation
time scale and the collision time scale.
For higher $\omega$ at the same chiral disk density $\rho$,
the peak in
$\theta_{\rm Hall}$
shifts to higher  
values of $F_{D}$.
We can compare these results to the behavior 
of $\theta_{\rm Hall}$ for driven skyrmions \cite{44,46,47}.  
In the absence of pinning, $\theta_{\rm Hall}$ for the skyrmion system has
a constant value determined by the materials properties \cite{44}.
When pinning is present, $\theta_{\rm Hall}$ gradually increases from zero at small
$F_D$,
similar
to what appears in Fig.~\ref{fig:2}(b).
In the skyrmion case, $\theta_{\rm Hall}$ saturates to the intrinsic value
at large $F_D$,
while for the chiral liquid, $\theta_{\rm Hall}$ decreases as $F_D$ increases above the
peak value.

\begin{figure}
\includegraphics[width=\columnwidth]{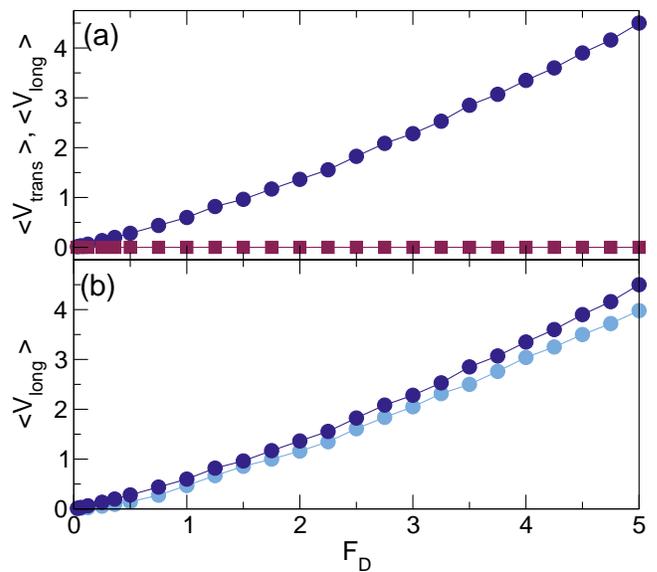}
\caption{ (a) $\langle V_{\rm long}\rangle$ (dark blue circles) and
  $\langle V_{\rm trans}\rangle$ (red squares) vs $F_{D}$ for a non-chiral fluid with the
  same parameters as in Fig.~\ref{fig:2} except with $A=0$.
  Here $\langle V_{\rm trans}\rangle = 0$ for all values of $F_{D}$.  
  (b) $\langle V_{\rm long}\rangle$ (dark blue circles) vs
  $F_{D}$ for the system in panel (a) at $A = 0.0$ and
  $\langle V_{\rm long}\rangle$ (light blue circles) vs $F_{D}$
  for the chiral system in Fig.~\ref{fig:2} with $A=2.5$, showing that the 
damping of the longitudinal motion is larger in the chiral fluid. 
}
\label{fig:4}
\end{figure}

In Fig.~\ref{fig:4}(a) we plot
$\langle V_{\rm long}\rangle$ and $\langle V_{\rm trans}\rangle$ versus $F_D$
for a non-chiral fluid with the same parameters as in Fig.~\ref{fig:2} but with
$A=0$.
Here, $\langle V_{\rm long}\rangle$ increases monotonically with increasing $F_D$,
similar to the chiral system;
however, $\langle V_{\rm trans}\rangle = 0$ for all values of $F_D$, indicating
that 
$\theta_{\rm Hall} = 0$
and that it is the
chiral motion of the bath particles that produces the Hall effect.
We find that
$\langle V_{\rm long}\rangle$
is slightly lower in the chiral liquid than for the $A=0$ passive disks,
as shown in 
Fig.~\ref{fig:4}(b) where we compare the $\langle V_{\rm long}\rangle$ versus $F_D$
curves for the $A=0$ system from Fig.~\ref{fig:4}(a) and the
$A=2.5$ system from Fig.~\ref{fig:2}(a).
The $A=2.5$ curve is lower for all $F_D$, indicating that the chirality of the bath
particles
increases the longitudinal drag on the probe particle.

\begin{figure}
\includegraphics[width=\columnwidth]{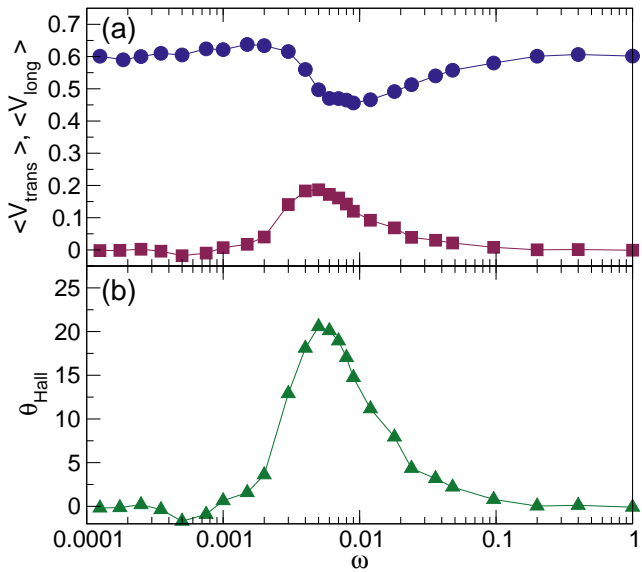}
  \caption{(a) $\langle V_{\rm trans}\rangle$ (red squares)
    and $\langle V_{\rm long}\rangle$ (blue circles)
    vs $\omega$ for a system with $F_{D} = 1.0$, 
    $A = 2.5$ and $\phi = 0.424$.
    A minimum in $\langle V_{\rm long}\rangle$ coincides with a maximum in
    $\langle V_{\rm trans}\rangle$ near $\omega  = 0.008$.
    (b) The corresponding $\theta_{\rm Hall}$ vs $\omega$
    showing a maximum Hall angle of $\theta_{\rm Hall}=23^{\circ}$ near 
$\omega = 0.004$.     
}
\label{fig:5}
\end{figure}

In Fig.~\ref{fig:5}(a) we plot $\langle V_{\rm long}\rangle$ and $\langle V_{\rm trans}\rangle$
versus $\omega$ for a system with $\phi = 0.424$, 
$A = 2.5$, and $F_{D} = 1.0$.
Here there is a dip in $\langle V_{\rm long}\rangle$ near $\omega=0.008$
that coincides with a peak in $\langle V_{\rm trans}\rangle$.
The corresponding $\theta_{\rm Hall}$ versus $\omega$ appears in Fig.~\ref{fig:5}(b),
where the Hall angle reaches a maximum value of
$\theta_{\rm Hall}=23^{\circ}$ near $\omega=0.004$.
At low frequencies, the chiral disks are rotating so slowly that the response
is close to that of a passive fluid,
while at high frequencies,
the chiral orbits diminish in radius
and the system again behaves like a passive fluid.

\begin{figure}
\includegraphics[width=\columnwidth]{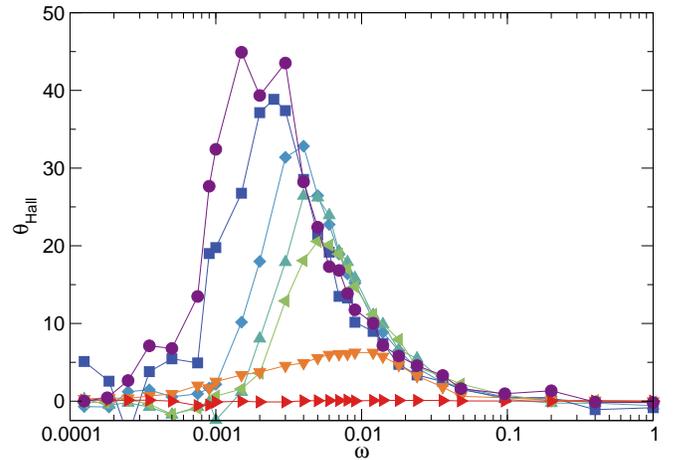}
\caption{
  $\theta_{\rm Hall}$ vs $\omega$ for the system in
  Fig.~\ref{fig:5}  with $A = 2.5$ and $\phi = 0.424$ for 
  $F_{D}=0.125$ (violet circles), $0.25$ (dark blue squares),
  $0.5$ (light blue diamonds), $0.75$ (teal up triangles),
  $1.0$ (green left triangles), $2.0$ (orange down triangles) and
  $4.0$ (red right triangles). 
  Here the maximum in $\theta_{\rm Hall}$ shifts to higher
  $\omega$ with increasing $F_{D}$ while the
  maximum possible Hall angle decreases.
}
\label{fig:6}
\end{figure}

Figure~\ref{fig:6}
shows $\theta_{\rm Hall}$ versus
$\omega$ for the system in Fig.~\ref{fig:5}
at $F_{D}$ values ranging from
$0.125$ to $4.0$.
The peak in $\theta_{\rm Hall}$ shifts to higher values of 
$\omega$ with increasing $F_{D}$
since the chiral particles must rotate faster in order to meet the resonance
condition $F_D\tau_P \sim a$ as $F_D$ increases.
The maximum value of $\theta_{\rm Hall}$ increases with decreasing $F_D$ since
the lower longitudinal velocity of the probe particle at the peak in $\theta_{\rm Hall}$
produces a longer collision time and thus a greater transfer of momentum from
the chiral disks
to the probe particle.
The maximum Hall angle we observe at very low
$F_{D}$ is close to $\theta_{\rm Hall} = 45^\circ$.

\begin{figure}
\includegraphics[width=\columnwidth]{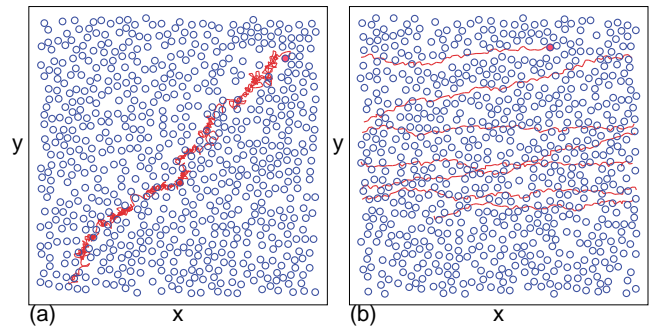}
\caption{Instantaneous positions of chiral disks (dark blue circles) and probe particle
  (red circle) along with the probe particle trajectory (red line) over a period
  of time for the system in Fig.~\ref{fig:6} with $A=2.5$ and $\phi=0.424$.
  (a) $F_{D} = 0.25$ and $\omega = 0.003$,
  where $\theta_{\rm Hall} \approx 45^\circ$.
(b) $F_{D} = 2.0$ and $\omega = 0.012$, where  $\theta_{\rm Hall} = 6.5^{\circ}$.  
}
\label{fig:7}
\end{figure}

In Fig.~\ref{fig:7}(a)
we plot the trajectory of the probe particle and the positions of the chiral bath particles
for the system in Fig.~\ref{fig:6}
at $F_{D} = 0.25$ and $\omega = 0.003$ where $\theta_{\rm Hall} \approx 45^\circ$.
During some time intervals, the probe particle moves at an angle of nearly
$90^\circ$ with respect to the driving direction. 
Figure~\ref{fig:7}(b) illustrates the same sample at 
$F_{D} = 2.0$ and $\omega = 0.012$, 
where the Hall angle is much smaller, $\theta_{\rm Hall} = 6.5^{\circ}$.

\begin{figure}
\includegraphics[width=\columnwidth]{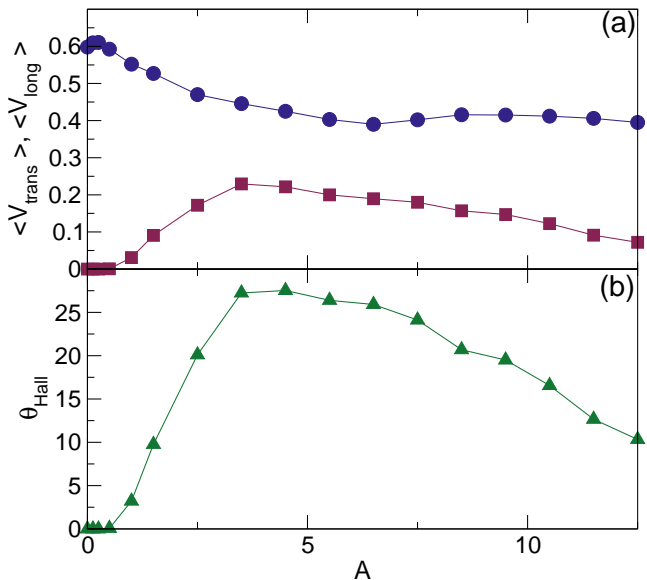}
\caption{(a) $\langle V_{\rm long}\rangle$ (blue circles) and $\langle V_{\rm trans}\rangle$
  (red squares) vs $A$
  for a system with $\omega = 0.006$, $F_{D} = 1.0$, and $\phi = 0.424$.
(b) The corresponding $\theta_{\rm Hall}$ vs $A$.
  There is a threshold value of $A_{c} = 0.5$ below which
  $\theta_{\rm Hall}=0$ and the Hall response is
  absent.
}
\label{fig:8}
\end{figure} 

In Fig.~\ref{fig:8}(a) we show $\langle V_{\rm trans}\rangle$ and
$\langle V_{\rm long}\rangle$ versus $A$ for a system with $\omega = 0.006$,
$F_{D} = 1.0$ and $\phi = 0.424$.
Here $\langle V_{\rm long}\rangle$ is large
in the $A = 0$ passive limit,
and it decreases with increasing $A$,
passing through a local minimum 
near $A = 7.0$.
We find that
there is a threshold value $A_{c}  = 0.5$
below  which $\langle V_{\rm trans}\rangle=0$ and there is no transverse response,
while a local maximum in $\langle V_{\rm trans}\rangle$
appears at $A = 4.0$.
We plot the corresponding $\theta_{\rm Hall}$ versus $A$
in Fig.~\ref{fig:8}(b), where the maximum value of
$\theta_{\rm Hall} = 27^{\circ}$ occurs    
near $A = 4.0$.

\begin{figure}
\includegraphics[width=\columnwidth]{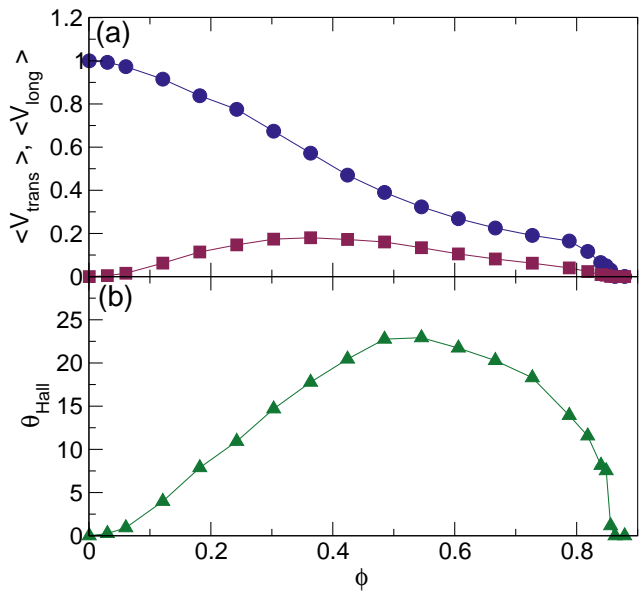}
\caption{
(a) $\langle V_{\rm long}\rangle$ (blue circles) and $\langle V_{\rm trans}\rangle$ 
  (red squares) vs $\phi$ for a system with
  $\omega = 0.006$, $F_{D} = 1.0$ and $A = 2.5$.
(b) The corresponding $\theta_{\rm Hall}$  vs $\phi$.  A jamming transition occurs
near $\phi = 0.86$.  
}
\label{fig:9}
\end{figure}

In Fig.~\ref{fig:9}(a) we plot $\langle V_{\rm long}\rangle$
and $\langle V_{\rm trans}\rangle$ versus $\phi$ for a system with 
$A = 2.5$, $F_{D} = 1.0$ and 
$\omega = 0.006$,
while in Fig.~\ref{fig:9}(b) we plot the corresponding
$\theta_{\rm Hall}$ versus $\phi$.
At the lowest densities, 
the probe particle undergoes
few collisions and moves in
the free flow limit
with $\langle V_{\rm long}\rangle/F_D = 1.0$ and $\langle V_{\rm trans}\rangle=0$.
As $\phi$ increases,
the probe particle velocity gradually decreases, dropping to zero near
$\phi = 0.86$ which is
the effective jamming density for these parameters. 
The decrease in the mobility of the probe particle with increasing density
and the vanishing of the mobility as
a jamming or crystallization density is approached resembles what was found
in previous studies of
active rheology for non-chiral
passive disk systems \cite{21,22,26,27}.
In those studies, $\langle V_{\rm trans}\rangle = 0$ for all values of $\phi$;
however, for the chiral disks we find
an increase in $\langle V_{\rm trans}\rangle$ with increasing density at low values
of $\phi$, with a maximum in
$\langle V_{\rm trans}\rangle$ appearing near $\phi=0.35$.
This low density increase in the transverse response results from the increasing
frequency of collisions between the probe particle and the chiral disks, since
it is these collisions that are responsible for the transverse probe particle motion.
For $\phi > 0.35$,
$\langle V_{\rm trans}\rangle$
decreases with increasing density
due to a crowding effect,
and at jamming
$\langle V_{\rm trans}\rangle$ drops to zero.
The maximum value of $\theta_{\rm Hall}$
occurs at a higher density of $\phi = 0.55$.

\begin{figure}
\includegraphics[width=\columnwidth]{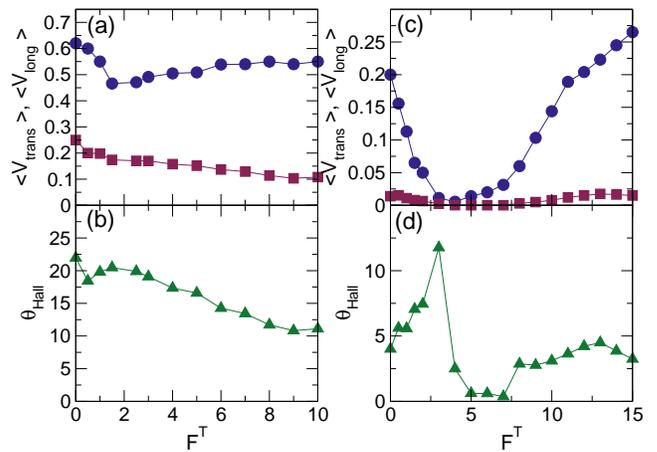}
\caption{
(a) $\langle V_{\rm long}\rangle$ (blue circles) and $\langle V_{\rm trans}\rangle$
  (red squares) vs $F^{T}$
  for a system with $\omega = 0.006$, $F_{D} = 1.0$, $A = 2.5$, and $\phi = 0.424$. 
(b) The corresponding $\theta_{\rm Hall}$ vs $F_{T}$.
(c) $\langle V_{\rm long}\rangle$ (blue circles) and $\langle V_{\rm trans}\rangle$
  (red squares) vs $F^{T}$ for a system with
  $\omega = 0.006$, $F_{D} = 1.0$, $A = 2.5$, and $\phi = 0.8482$.
  (d) The corresponding $\theta_{\rm Hall}$ vs $F_{T}$.
  Here we find
  a freezing by heating phenomenon 
in the interval $4.0 < F^{T} < 7.0$.   
}
\label{fig:10}
\end{figure}

\begin{figure}
\includegraphics[width=\columnwidth]{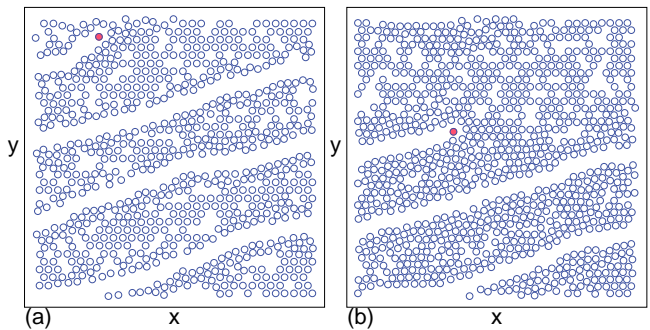}
\caption{Instantaneous positions of chiral disks (dark blue circles) and probe particle
  (red circle) for the system in Fig.~\ref{fig:10}(a)
  with $\omega=0.006$, $F_D=1.0$, $A=2.5$, and $\phi=0.424$
  at $F^{T} = 0$, where the
  probe particle has a finite Hall angle but
  leaves a low density 
  depletion zone or wake behind as it moves.
  (b) The same system at a higher density of $\phi = 0.67$. Here the probe particle mobility
  is reduced but the Hall angle remains finite and the depletion zone persists.}
\label{fig:11}
\end{figure}

We next consider the effect of changing the
magnitude of the thermal fluctuations.
In Fig.~\ref{fig:10}(a) we plot $\langle V_{\rm long}\rangle$
and $\langle V_{\rm trans}\rangle$
versus $F^T$ for a system with $F_{D} = 1.0$, $\omega = 0.006$, 
$A = 2.5$, and $\phi = 0.424$.
At $F^{T} = 0$, when the system is in the granular limit, the
probe  particle leaves a low density wake behind it
and 
$\langle V_{\rm trans}\rangle$ remains finite,
indicating that thermal fluctuations are not 
necessary to produce the Hall response.
In Fig.~\ref{fig:10}(a), $\langle V_{\rm trans}\rangle$ monotonically decreases  
with increasing $F^{T}$; however, there is a local minimum in
$\langle V_{\rm long}\rangle$ near $F^{T} = 2.0$.
The local minimum roughly coincides with
the crossover between low temperatures, where the probe leaves behind a low
density wake, and higher temperatures, where the wake rapidly refills with chiral
disks.
Figure~\ref{fig:10}(b) shows that the corresponding $\theta_{\rm Hall}$
versus $F^T$ has its
maximum value at $F^{T} = 0$,
with a smaller local maximum appearing near
$F^{T} = 2.0$.  Above $F^T=2.0$, $\theta_{\rm Hall}$ decreases monotonically
with increasing $F^T$.

In Fig.~\ref{fig:11}(a) we illustrate the probe particle motion
for the system in Fig.~\ref{fig:10}(a) with $\phi=0.424$ at $F^{T} = 0$, where
the probe particle moves at a finite Hall angle and
leaves a low density wake in its path.
The appearance of an
empty region behind the probe particle is similar to what has been observed for active 
rheology of non-thermal granular 
materials below the jamming density \cite{22,27,28}, since in these systems there is
no energy penalty for the formation of a void.
At finite $F^{T}$, the 
chiral disks diffusively fill in the empty space.
In Fig.~\ref{fig:11}(b) we show the probe particle motion over the same time interval
in a denser system with $F^T=0$ and $\phi=0.67$.
The probe particle does not
translate as far due to the decrease in mobility;
however, it still leaves behind
a low density wake.

In Fig.~\ref{fig:10}(c) we plot $\langle V_{\rm long}\rangle$
and $\langle V_{\rm trans}\rangle$ versus $F^{T}$ 
for a high density system with $\phi = 0.8482$,
$\omega = 0.006$, $A = 2.5$, and $F_{D} = 1.0$,
and in Fig.~\ref{fig:10}(d) we show the corresponding $\theta_{\rm Hall}$ 
versus $F^{T}$.
These parameters fall within
a low mobility regime, where the probe particle is not stuck but can only move
relatively slowly through the chiral bath.
At $F^{T} = 0$, $\langle V_{\rm long}\rangle \approx 0.2$
and $\theta_{\rm Hall} = 4.5^\circ$.
As $F^{T}$ increases, both $\langle V_{\rm long}\rangle$
and $\langle V_{\rm trans}\rangle$ decrease, reaching a value that is
close to zero near $F^{T} = 4.0$.
This is a signature of a thermally-induced jamming transition that occurs
when the thermal fluctuations increase the effective size of the bath particles
and raise the effective density of the system to the jamming density.
Such a transition can also be regarded as an example of a
freezing by heating 
phenomenon
in which the thermal fluctuations can effectively freeze the
disks into a jammed state \cite{49}. 
If the thermal fluctuations are finite but small,
the chiral disks maintain their ordering in the jammed state
and the probe 
particles slowly makes its way through
the resulting mostly triangular solid.
As $F^{T}$ increases, the fluctuations become
strong enough to melt the chiral disk crystal.
As a result, liquid behavior reappears
and the probe particle mobility rebounds,
leading to the increase
in both $\langle V_{\rm long}\rangle$ and
$\langle V_{\rm trans}\rangle$
for $F^{T}> 6.0$.
The Hall angle $\theta_{\rm Hall}$ in Fig.~\ref{fig:10}(d)
passes through a local maximum near $F^{T} = 2.5$ just before
the onset of thermally-induced jamming, and it drops nearly to zero
within the jammed state when the probe particle motion becomes
extremely slow.
For $F^{T} > 6.5$, when the jammed state melts and the probe particle
mobility increases,
$\theta_{\rm Hall}$ increases back to its pre-jammed level.
These results
indicate that a finite Hall effect
can be observed even in
non-thermal chiral systems. 

\begin{figure}
\includegraphics[width=\columnwidth]{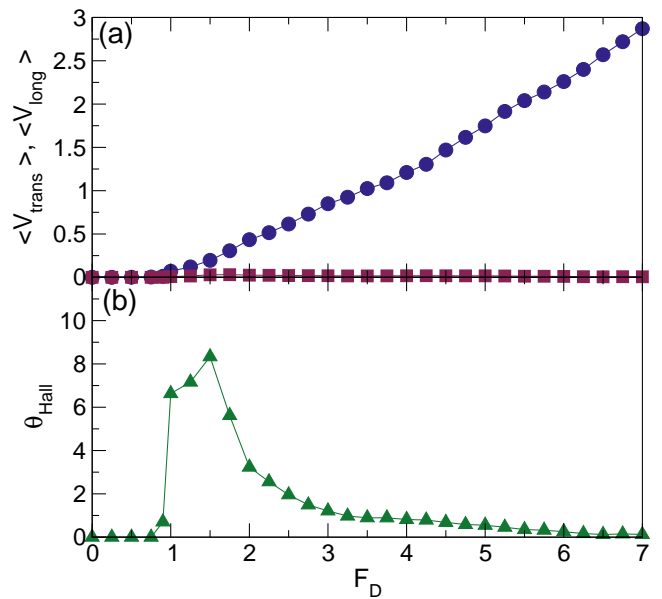}
\caption{(a) $\langle V_{\rm long}\rangle$ (blue circles)
  and $\langle V_{\rm trans}\rangle$ (red squares) vs $F_{D}$ 
  for a system with $\phi = 0.848$,
  $A = 2.5$, and $F^{T} = 2.0$.
  There is a finite depinning threshold
  near $F_{D} = 0.6$.
  (b) The corresponding $\theta_{\rm Hall}$ vs $F_{D}$ passes through a maximum
at $F_{D} = 1.75$. 
}
\label{fig:12}
\end{figure}

Near the jammed state at high chiral disk densities,  the behavior of
$\theta_{\rm Hall}$ depends strongly on $F_{D}$ and $\omega$,
since
the probe particle can only move through the jammed chiral disks
if the driving force is larger than a depinning threshold $F_c$.
A monodisperse assembly of passive disks at $T = 0$
forms a triangular solid at a density of $\phi = 0.9$.
For densities close to but below this solidification density,
the addition of thermal fluctuations can
induce freezing by heating or the formation of grain boundaries and other
defects, and the disks exhibit
glassy or very slow dynamics for densities in the range
$0.83 < \phi < 0.9$. 
In our chiral disk system at $F^T=2.0$ and $\phi=0.8482$, the probe particle is mobile
when $F_{D} = 1.0$,
but if we reduce $F_D$ we find that there is
a finite threshold drive $F_c$ below which the probe particle is no longer able to move.
This is illustrated in
Fig.~\ref{fig:12}(a), where we plot $\langle V_{\rm long}\rangle$
and $\langle V_{\rm trans}\rangle$ versus $F_{D}$ 
for a system with $\phi = 0.8482$, $\omega = 0.006$, $F^T=2.0$, and $A = 2.5$.
Here $\langle V_{\rm long}\rangle = \langle V_{\rm trans}\rangle = 0$
when $F_D<F_c$, where the threshold force
$F_{c} = 0.6$.
In the corresponding $\theta_{\rm Hall}$ versus $F_{D}$ curve shown
in
Fig.~\ref{fig:12}(b),
we find that
$\theta_{\rm Hall} = 0$ for $F_{D} < 0.8$,
indicating that within the window $F_c < F_D < 0.8$,
the probe particle has a finite longitudinal velocity but exhibits no Hall effect.
The Hall angle reaches its maximum value of
$\theta_{\rm Hall} \approx 8.5$
near $F_{D} = 1.75$,
and gradually decreases toward zero for higher drives.
Since this system is at a finite temperature of $F^T=2.0$,
the probe particle is best described as undergoing a creep behavior
at $F_{D} = 0.8$.
During long periods of time, the probe particle is pinned, but there are occasional
events in which the probe particle jumps to a new pinned location.
Recent studies of driven skyrmions \cite{46,50} showed that the Hall
angle is zero
in the creep
regime
and becomes finite at higher drives when the skyrmions transition to continuous
flow, similar
to what we observe in Fig.~\ref{fig:12}; however, in the skyrmion case,
$\theta_{\rm Hall}$ saturates to the intrinsic value at high drives
rather than decreasing back to zero as in Fig.~\ref{fig:12}.

\begin{figure}
\includegraphics[width=\columnwidth]{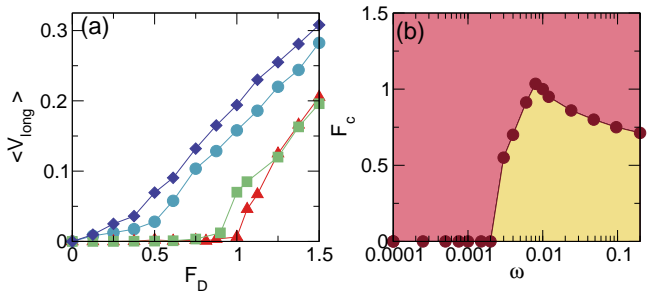}
\caption{(a) $\langle V_{\rm long}\rangle$ vs $F_{D}$
  for a system with $\phi = 0.8482$, $A = 2.5$, and $F^{T}  = 2.0$ at
  $\omega = 0.008$ (red triangles), $0.006$ (green squares),
  $0.003$ (light blue circles), and $0.001$ (dark blue diamonds). 
  (b) The depinning force $F_{c}$ vs $\omega$ for the same system
  highlighting regions in which the
  probe particle is moving (pink) or pinned (yellow).
}
\label{fig:13}
\end{figure}

At high densities,
we find that the
threshold force $F_{c}$ depends on the frequency at which the chiral disks rotate.
In Fig.~\ref{fig:13}(a) we plot
$\langle V_{\rm long}\rangle$ versus
$F_{D}$ in a system with $A = 2.5$, $F^{T} = 2.0$, and $\phi = 0.8482$ at
$\omega = 0.008$, $0.006$, $0.003$, and
$0.001$.
The threshold for motion is finite
at $\omega = 0.006$ and $\omega=0.008$,
and zero for $\omega=0.003$ and $\omega=0.001$. 
In Fig.~\ref{fig:13}(b) we show $F_{c}$ versus $\omega$
for the system from Fig.~\ref{fig:13}(a).  For drives above
$F_c$, the probe can flow through the sample, but for drives below $F_c$, the
probe particle is pinned.
Here
we find that $F_c$ is finite only when
$\omega > 0.003$, and that there is a
local maximum in $F_{c}$ near $\omega = 0.01$.
These results indicate that at higher disk densities,
the activity level of the chiral disks can be used to    
control a transition from jammed to unjammed behavior.

\section{Summary} 

We have numerically examined the motion of a probe particle 
driven through a chiral liquid composed of circularly moving disks.
In the absence of chirality, the
probe particle drifts only in the direction of drive so there is
no Hall effect; however, when
the bath particles are chiral,
both the longitudinal and transverse velocities of the probe particle are finite.
Since a portion  
of the probe particle motion is perpendicular to the drive direction,
the probe particle exhibits a
finite Hall angle
similar to what is observed for
a charged particle moving in a magnetic field or for driven skyrmion systems.
We find that
the Hall angle has a 
non-monotonic
dependence on the probe particle driving force
and the amplitude and frequency of the chiral disk motion.
At low drives, the probe particle can undergo multiple collisions
with an individual chiral bath particle, reducing the Hall angle,
while at high drives the collisions between the probe and chiral bath particles are very
brief, which again reduces the magnitude of the Hall angle.
An optimal Hall angle occurs
when the time between collisions of the probe particle with consecutive
chiral bath particles
is roughly
equal to the time required
for a chiral bath particle to complete a single rotation.
We find that
the Hall angle can reach values as large as
$\theta_{\rm Hall}=45^{\circ}$
and that the Hall effect persists in the zero temperature or granular limit. 
When the chiral disk activity is fixed, the Hall angle is maximized at an
optimal chiral disk density,
while
the probe particle velocity in
both the longitudinal and transverse directions drops to zero when the
chiral disks reach the jamming density,
which is dependent on
the frequency of the
chiral motion.
At low frequencies, the depinning threshold is zero and the probe particle
is able to move under all applied drives,
while at higher frequencies there is a finite depinning threshold, and for drives
below this threshold, the probe particle is pinned.
We compare our results with those obtained for
skyrmions moving over random disorder,
where drive-dependent Hall angles that increase with increasing drive
are observed.
In the skyrmion case, the
Hall angle saturates to the clean limit at high drives, whereas for the chiral liquid
we consider here, the Hall angle decreases to zero at high drives.

{\it Note added--} In the course of completing this work,
we became aware of the work of
Kumar {\it et al.} \cite{51} on the motion of spinning probe particles through granular
matter, 
where they report the onset of a Magnus like effect including a lift force.   

\begin{acknowledgments}
We gratefully acknowledge the support of the U.S. Department of
Energy through the LANL/LDRD program for this work.
This work was supported by the US Department of Energy through
the Los Alamos National Laboratory.  Los Alamos National Laboratory is
operated by Triad National Security, LLC, for the National Nuclear Security
Administration of the U. S. Department of Energy (Contract No. 892333218NCA000001).
\end{acknowledgments}

\end{document}